\documentclass[prd,twocolumn,showpacs,showkeys,preprintnumbers,floatfix,nofootinbib]{revtex4-1}

\usepackage{amsfonts} 
\usepackage{amssymb} 
\usepackage{amsmath} 
\usepackage{subfigure}
\usepackage{graphicx} 
\usepackage{array} 
\usepackage{dcolumn} 
\usepackage{bm} 
\usepackage{latexsym} 
\usepackage{longtable} 
\usepackage{hyperref} 
\usepackage{color}
\usepackage{colortbl}
\usepackage[usenames,dvipsnames]{xcolor}
\usepackage{mathrsfs}
\usepackage{comment}
\usepackage{rotating}
\usepackage{here}
\usepackage{float}
\usepackage{multirow}
\usepackage{hyperref}

\definecolor{green}{rgb}{0.0,0.6,0.0}
\definecolor{blue}{rgb}{0.0,0.0,0.6}

\graphicspath{{./figs/}}

\allowdisplaybreaks

\providecommand{\wbar}[1]{\overline#1}
\providecommand{\abs}[1]{\lvert#1\rvert}

\providecommand{\expv}[1]{\langle#1\rangle}

\providecommand{\CL}{\nonumber\\}

\newcolumntype{C}[1]{>{\centering\arraybackslash}p{#1}}
\newcolumntype{.}[1]{D{.}{.}{#1}}

\definecolor{HLBlue}{HTML}{6599FF}
\definecolor{HLOrange}{HTML}{FF6600}

\usepackage{bm}

\newcommand{\Dslash}{\ensuremath{{D\kern -0.65em /}}}

\newcommand{\sign}{\mathop{\mathrm{sgn}}}

\newcommand{\order}{\ensuremath{\mathrm{O}}}

\begin{document}

\title{Heavy-quark meson spectrum tests of the Oktay-Kronfeld action}

\author{Jon A. Bailey}
\affiliation{
  Lattice Gauge Theory Research Center, FPRD, and CTP,
  Department of Physics and Astronomy, \\
  Seoul National University, Seoul, 151-747, South Korea
}

\author{Carleton DeTar}
\email[E-mail: ]{detar@physics.utah.edu}
\affiliation{
  Department of Physics and Astronomy, University of Utah,
  Salt Lake City, UT  84112, USA
}

\author{Yong-Chull Jang}
\email[E-mail: ]{ypj@lanl.gov}
\altaffiliation{present address: Theoretical Division T-2, Los Alamos National Laboratory, Los Alamos, NM 87545, USA.}
\affiliation{
  Lattice Gauge Theory Research Center, FPRD, and CTP,
  Department of Physics and Astronomy, \\
  Seoul National University, Seoul, 151-747, South Korea
}

\author{Andreas S. Kronfeld}
\email[E-mail: ]{ask@fnal.gov}
\affiliation{
  Theoretical Physics Department, Fermi National Accelerator Laboratory, Batavia, IL  60510, USA \\
  Institute for Advanced Study, Technische Universit\"at M\"unchen, 85748 Garching, Germany}

\author{Weonjong Lee}
\email[E-mail: ]{wlee@snu.ac.kr}
\affiliation{
  Lattice Gauge Theory Research Center, FPRD, and CTP,
  Department of Physics and Astronomy, \\
  Seoul National University, Seoul, 151-747, South Korea
}

\author{Mehmet B. Oktay}
\affiliation{
  Department of Physics and Astronomy, University of Utah, Salt Lake City, UT  84112, USA \\
  Department of Physics and Astronomy, University of Iowa, Iowa City, IA 52242, USA 
}

\collaboration{SWME, Fermilab Lattice, and MILC Collaborations}

\date{\today}

\begin{abstract}
The Oktay-Kronfeld (OK) action extends the Fermilab improvement program for massive Wilson fermions to
higher order in suitable power-counting schemes.
It includes dimension-six and -seven operators necessary for matching to QCD through order
$\order(\Lambda^3/m_Q^3)$ in HQET power counting, for applications to heavy-light systems, and $\order(v^6)$
in NRQCD power counting, for applications to quarkonia.
In the Symanzik power counting of lattice gauge theory near the continuum limit, the OK action includes all
$\order(a^2)$ and some $\order(a^3)$ terms.
To assess whether the theoretical improvement is realized in practice, we study combinations of
heavy-strange and quarkonia masses and mass splittings, designed to isolate heavy-quark discretization
effects.
We find that, with one exception, the results obtained with the tree-level-matched OK action are
significantly closer to the continuum limit than the results obtained with the Fermilab action.
The exception is the hyperfine splitting of the bottom-strange system, for which our statistical errors are
too large to draw a firm conclusion.
These studies are carried out with data generated with the tadpole-improved Fermilab and OK actions on 500
gauge configurations from one of MILC's $a\approx0.12$~fm, $N_f=2+1$-flavor, asqtad-staggered ensembles.
\end{abstract}

\maketitle

\section{Introduction}
\label{sec:intro}

Lattice QCD has been used to calculate nonperturbatively hadronic matrix elements needed for many aspects of
Standard Model (SM) phenomenology.
For example, hadronic matrix elements of flavor-changing currents are crucial to quark-flavor physics (see,
e.g., Ref.~\cite{Du:2015tda}), and many of them are gold-plated calculations~\cite{Davies:2003ik} in 
lattice~QCD.
For gold-plated light-quark processes, lattice QCD has achieved high precision~\cite{Aoki:2016frl}.
Especially notable is the hadronic matrix element $B_K$ needed to understand indirect CP violation in the
neutral kaon system, which is known to 1.3\% \cite{Durr:2011ap,Laiho:2011np,Blum:2014tka,Carrasco:2015pra,%
Jang:2015sla,Aoki:2016frl}.
That said, simulating heavy quarks in lattice QCD is a long-standing challenge~\cite{Kronfeld:2003sd,%
El-Khadra:hep-lat/1403.5252,Bouchard:hep-lat/1501.03204}, and in this paper we
examine a way to control discretization effects for massive (i.e., $m_Q\gg\Lambda$) fermions in lattice 
gauge theory~\cite{EKM,Kronfeld:PhysRevD.62.014505,Oktay2008:PhysRevD.78.014504,Burch:2009az}.

In the SM, the Cabibbo-Kobayashi-Maskawa (CKM) matrix explains all changes of quark flavor.
The CKM matrix contains four independent parameters, three angles and one phase, with the phase being
responsible for CP violation.
Three of the four parameters are the magnitude and phase of the matrix element $V_{ub}$ and the magnitude of
$V_{cb}$.
In practice, $|V_{cb}|$, $|V_{ub}|$, and $\arg V_{ub}$ are determined in $B$-meson decays and CP asymmetries.
The remaining CKM parameter is the Cabibbo angle, i.e., $|V_{us}|$, which is known from kaon decays to 
0.3\%~\cite{Bazavov:2013maa,Bazavov:2014wgs,Olive:2016xmw}.

In $B$ physics, the unitarity of the CKM matrix is often visualized with the unitarity triangle (UT).
Indirect CP violation in the neutral kaon system constrains the apex of the UT via the quantity
$\varepsilon_K$.
In the SM, $\varepsilon_K$ depends parametrically on $|V_{cb}|$ to the fourth power.
There is a $3.2\sigma$ tension between $\varepsilon_K$ in the SM~\cite{Bailey:2016dzk,Lee:2016xkb,%
Bailey2015:PhysRevD.epsK} and experiment~\cite{Ambrosino:2006ek} when the exclusive value of
$|V_{cb}|$~\cite{DeTar:2015orc} from the decays $\bar{B}\to D^{(*)}\ell\bar{\nu}$ is taken as input.
In contrast with the 0.5\% uncertainty in the experimental value of $\varepsilon_K$, there is a 10\%
uncertainty in this SM prediction, mostly coming from an uncertainty in the CKM matrix element~$|V_{cb}|$.

In addition, there is a $3\sigma$ tension between the exclusive~\cite{Bailey2014:PhysRevD.89.114504} and
inclusive~\cite{Alberti2014:PhysRevLett.114.061802} determinations of $|V_{cb}|$, and a similar tension of
2--3$\sigma$ in $|V_{ub}|$ \cite{Ricciardi:hep-ph/1412.4288, Ricciardi:ModPhysLettA.28.1330016}.
Therefore, to conduct stringent tests of the SM description of flavor physics, reducing the errors of
$|V_{ub}|$ and $|V_{cb}|$ is necessary.
At present, the experimental and theoretical errors on $|V_{ub}|$ and $|V_{cb}|$ are comparable.
The target is promising because the experimental precision of the decay rates will get better with the
Belle~II and LHCb experiments.
On the theoretical side, the errors in the form factors of the semileptonic decays, %
$\bar{B}\to D^{(*)}\ell\bar{\nu}$, for $|V_{cb}|$, and $\bar{B}\to\pi\ell\bar{\nu}$, for $|V_{ub}|$, must be
reduced.

Another interesting quantity is the $B_{(s)}$-meson decay constant, $f_{B_{(s)}}$. 
Its physical significance is highlighted in the rare leptonic decay $B_s \to \mu^+\mu^-$.
This decay is sensitive to new physics, because, as a flavor-changing neutral-current process, it is a
suppressed, higher-order loop-effect in the SM.
Thus, the decay $B_s \to \mu^+\mu^-$ can place strong constraints on beyond the SM models
\cite{El-Khadra:hep-lat/1403.5252,Horgan2013:PhysRevLett.112.212003,Horgan2015:hep-lat/1501.00367}.
Currently, many lattice-QCD collaborations have calculated the decay constant
$f_{B_{(s)}}$~\cite{Bazavov2011:PhysRevD.85.114506,McNeile2011:PhysRevD.85.031503,%
Na2012:PhysRevD.86.034506,Dowdall2013:PhysRevLett.110.222003,Christ2014:PhysRevD.91.054502}, and
the results are consistent within errors~\cite{Aoki:2016frl}.
Increasing the precision of the lattice-QCD determination will provide stronger constraints.

A~major challenge in increasing the precision of lattice QCD calculations of heavy-quark quantities is to
control heavy-quark discretization errors.
Because the heavy-quark, $b$ and $c$, masses and the accessible ultraviolet cutoff $a^{-1}$ are comparable,
special care is needed to handle heavy quark discretization effects.
Following the development of nonrelativistic QED for the study of bound
states~\cite{Caswell:PhysLettB.167.437}, nonrelativistic QCD (NRQCD) was developed for lattice calculations
of hadronic bound states~\cite{Lepage:NuclPhysB.4.199, Thacker:PhysRevD.43.196, Davies:PhysRevD.45.915,
Lepage:PhysRevD.46.4052} and quarkonium phenomenology~\cite{Bodwin:PhysRevD.51.1125}.
A~similar approach, heavy quark effective theory (HQET), was adopted for lattice calculations of heavy-light
systems, treating heavy quarks as a static chromoelectric source~\cite{Eichten:1987xu,%
Eichten:PhysLettB.234.511,Eichten:PhysLettB.240.193}.

Whereas NRQCD and HQET are nonrelativistic effective field theories treating heavy quarks as two-component
spinors, the Fermilab action~\cite{EKM} is based on the Wilson clover action~\cite{Wilson:1975id,%
Sheikholeslami:NuclPhysB.259.572} with four-component spinors.
To retain a suitable remnant of Lorentz symmetry in physical quantities, the lattice action must have
a time-space axis-interchange asymmetric form.
The Fermilab action is then tuned with explicitly (bare) mass dependent couplings $c_i(m_0a)$ to
connect smoothly the large $m_Qa>1$ and small $m_Qa\ll1$ mass limits.
In this way, the discretization errors can be controlled for arbitrary quark mass.
In addition, a continuum limit without fine tuning is possible with the Fermilab action, which is not the 
case with direct discretizations of NRQCD.

The Fermilab action is applicable to both heavy-light systems and quarkonia, with suitable power counting:
HQET power counting \cite{Kronfeld:PhysRevD.62.014505,Oktay2008:PhysRevD.78.014504} for heavy-light systems
and NRQCD for quarkonia \cite{Lepage:PhysRevD.46.4052, Oktay2008:PhysRevD.78.014504,Burch:2009az}.
Using ideas from the effective theories, one can develop a HQET- or NRQCD-based theory of cutoff
effects~\cite{EKM,Kronfeld:PhysRevD.62.014505,Oktay2008:PhysRevD.78.014504,Burch:2009az}, yielding a
systematic improvement and error estimation method.
This nonrelativistic analysis establishes a different interpretation, in which one tuning condition is not
imposed, without introducing large discretization effects into matrix elements and mass splittings.
This technique, often known as the Fermilab nonrelativistic interpretation, is used in the heavy-quark
calculations of the Fermilab Lattice and MILC Collaborations.
Related approaches~\cite{Aoki:2001ra,Christ2006:PhysRevD.76.074505} essentially use the Fermilab action, but
impose the tuning condition left out in the Fermilab nonrelativistic interpretation.

The Oktay-Kronfeld (OK) action~\cite{Oktay2008:PhysRevD.78.014504} is an extension of the Fermilab action
that incorporates all dimension-six and certain dimension-seven bilinear operators.
Although there are many such operators, many can be eliminated as redundant.
Furthermore, only six are needed for tree-level matching to QCD through $\order(\lambda^3)$
($\lambda\sim\Lambda_\text{QCD}/m_Q$ or $\Lambda_\text{QCD}a$) in HQET power counting for heavy-light mesons
and $\order(v^6)$ ($v$ being the relative velocity of the quark and antiquark) in NRQCD power counting for
quarkonium.
When $m_0a\ll1$, OK improvement reduces to $\order(a^2)$ with some $\order(a^3)$ terms in
Symanzik~\cite{Symanzik:NuclPhysB.226.187} power counting.
It is expected that the bottom- and charm-quark discretization errors could be reduced below the $1\%$~level
with the OK action~\cite{Oktay2008:PhysRevD.78.014504}.
A~similar error can be achieved with other highly improved actions, such as highly-improved staggered
quarks~\cite{Follana:2006rc}, once the lattice spacing is small enough.

To test whether the theoretical improvement~\cite{Oktay2008:PhysRevD.78.014504} is achieved in practice, we
form combinations of heavy-light meson and quarkonium masses, designed to isolate discretization errors.
For spin-independent improvements, we compute a combination of heavy-light and quarkonium masses discussed
in Refs.~\cite{Collins,Kronfeld}.
For spin-dependent improvements, we compute the hyperfine splittings in each system.
For the work reported here, we generate data using a MILC asqtad-staggered $N_f=2+1$ ensemble with
$a\approx0.12$~fm \cite{Bazavov:RevModPhys.82.1349}.
We use an optimized conjugate gradient inverter \cite{Jang:LAT2013} to calculate quark propagators with both
the tadpole-improved Fermilab and OK actions over ranges spanning the $b$ and $c$ quarks.
For the $b$-like and $c$-like quarks, the data are generated with four different values of the hopping
parameter for the OK action, and two for the Fermilab action.
Tadpole improvement of all terms is fully implemented in the conjugate-gradient program, as in an early,
preliminary study~\cite{MBO:LAT2010}.
A~preliminary report of the analysis reported here can be found in Ref.~\cite{Bailey2014:LAT2014.097}.

The outline of this paper is as follows.
In Sec.~\ref{sec:oka}, we briefly review the OK action and discuss its tadpole improvement.
In Sec.~\ref{sec:corr}, we describe the correlators we generate and the fits we use to extract the
lightest meson energies in each channel at each momentum.
In Sec.~\ref{sec:disp}, we describe the fits needed to extract the rest and kinetic masses of the
pseudoscalar and vector mesons.
In Sec.~\ref{sec:icp}, we assess the improvement from higher-order kinetic terms in the OK action with the
mass combination of Refs.~\cite{Collins,Kronfeld}.
In Sec.~\ref{sec:hfs}, we assess the improvement from higher-order chromomagnetic interactions in the OK
action by inspecting the difference of the hyperfine splittings of the rest and kinetic masses.
We conclude in Sec.~\ref{sec:con}.

\section{Oktay-Kronfeld Action}
\label{sec:oka}

The Oktay-Kronfeld (OK) action~\cite{Oktay2008:PhysRevD.78.014504} $S_\text{OK}$ is an improved version of
the Fermilab action~\cite{EKM} $S_\text{FL}$, which takes the clover
action~\cite{Sheikholeslami:NuclPhysB.259.572} for Wilson fermions~\cite{Wilson:1975id}, but chooses
time-space asymmetric couplings:
\begin{align}
    S_\text{FL} &= S_0 + S_B + S_E , \\
    S_\text{OK} &= S_\text{FL} + S_\text{new} .
\end{align}
The explicit form of each piece $S_0$, $S_B$, $S_E$, and $S_\text{new}$, including many redundant operators
and several operators not needed for tree-level matching, can be found in
Ref.~\cite{Oktay2008:PhysRevD.78.014504}.
Here, we give an appropriate form for numerical simulations with tree-level tadpole
improvement~\cite{Lepage:PhysRevD.48.2250}.
We write the action with gauge covariant translation operators
$T_{\pm\mu}\psi(x)=U_{\pm\mu}(x)\psi(x\pm a\hat{\mu})$ with $U_{-\mu}(x) = U^\dagger_\mu(x-a\hat{\mu})$.

We write the the action with dimensionless fields $E_{x,i}$, $B_{x,i}$, $\psi_x$, and $\bar{\psi}_x$,
\begin{align}
    E_{x,i} &= a^{2} E_{i}(x) , \\
    B_{x,i} &= a^{2} B_{i}(x) ,\\
    \psi_{x}       &= a^{3/2} \psi(x) , \\
    \bar{\psi}_{x} &= a^{3/2} \bar{\psi}(x) ,
\end{align}
where the chromomagnetic field $B_{i}(x)$ and chromoelectric field $E_{i}(x)$ are
\begin{align}
    B_{i}(x) &= \frac{1}{2} \sum_{j,k=1}^{3} \epsilon_{ijk} F_{jk}(x) ,\\
    E_{i}(x) &= F_{4i}(x) ,
\end{align}
and
\begin{equation}
    F_{\mu\nu} = \frac{1}{8a^2} \sum_{\rho = \pm \mu} \sum_{\sigma = \pm \nu} \sign(\rho\sigma)
        \left[ T_{\rho}T_{\sigma}T_{-\rho}T_{-\sigma} - \text{H.c.} \right] 
\end{equation}
is the four-leaf-clover field strength.

Once the action is written in terms of the translation operators $T_\mu$, tadpole improvement can be carried
out with the following steps~\cite{EKM}: %
Rewrite the gauge covariant translation operators $T_{\pm\mu}$ in terms of tadpole-improved ones
$\tilde{T}_{\pm\mu}$ by using $T_{\pm\mu}=u_0\tilde{T}_{\pm\mu}$; replace the couplings $c_i\to\tilde{c}_i$,
$r_s\to\tilde{r}_s$, $\zeta\to\tilde{\zeta}$, and $m_0\to\tilde{m}_0$ (the $u_0$ factors are absorbed into
the couplings $\tilde{c}_i$, $\tilde{r}_s$, $\tilde{\zeta}$, and $\tilde{m}_0$); and finally, multiply each
term of the action by the factor $u_0$ to recover the original form of the action.
Then, we arrive at the tadpole-improved OK action in terms of the unimproved
translation operators, explicit tadpole factors, the tadpole-improved couplings~$\tilde{c}_i$, and the
hopping parameter~$\kappa$:
\begin{align}
    \kappa &= \tilde{\kappa}/u_0  , \\
    \frac{1}{2\tilde{\kappa}} &= \tilde{m}_0 a + ( 1 + 3\tilde{r}_{s} \tilde{\zeta} + 18 \tilde{c}_4 ).
    \label{eq:kappa}
\end{align}
The OK action then is
\begin{align}
  S_\text{OK} 
  &= \frac{1}{2\kappa} \bar{\psi}_x \psi_{x} \CL
  &-\frac{1}{2} \bar{\psi}_x (1-\gamma_{4}) T_{4} \psi_{x}
  -\frac{1}{2} \bar{\psi}_x (1+\gamma_{4}) T_{-4}\psi_{x} \CL
  &-\frac{1}{2} \bar{\psi}_x \sum_{i=1}^{3} 
  ( \tilde{r}_{s}\tilde{\zeta} + 8\tilde{c}_{4} 
  -\gamma_{i}(\tilde{\zeta} - 2\tilde{c}_{1} - 12\tilde{c}_{2}) ) 
   T_{i} \psi_{x} \CL
   &-\frac{1}{2} \bar{\psi}_x \sum_{i=1}^{3} 
   ( \tilde{r}_{s}\tilde{\zeta} + 8\tilde{c}_{4} 
   +\gamma_{i}(\tilde{\zeta} - 2\tilde{c}_{1} - 12\tilde{c}_{2}) ) 
   T_{-i} \psi_{x} \CL
  &+\frac{(\tilde{c}_{1} + 2\tilde{c}_{2})}{2u_0}  
  \bar{\psi}_x \sum_{i=1}^{3} \gamma_{i} (T_{i}T_{i} - T_{-i}T_{-i}) \psi_{x} \CL
  &+\frac{\tilde{c}_{4}}{u_0}
  \bar{\psi}_x \sum_{i=1}^{3} (T_{i}T_{i} + T_{-i}T_{-i}) \psi_{x} \CL
  &+\frac{\tilde{c}_{2}}{2u_0}
  \bar{\psi}_x \sum_{i=1}^{3} \sum_{j \neq i}^{3} \gamma_{i} 
    \{ (T_{i}-T_{-i}), (T_{j}+T_{-j}) \} \psi_{x} \CL
  &+i\frac{\tilde{c}_{5}}{4 u_0^2}
  \bar{\psi}_x \sum_{i=1}^{3} \Sigma_{i} T^{(3)}_{i} \psi_{x} \CL
  &-\frac{\tilde{c}_E \tilde{\zeta}}{2u_0^3}
  \bar{\psi}_x \sum_{i=1}^{3} \alpha_{i} E_{x,i} \psi_{x} \CL
  &-i\frac{(\tilde{c}_B \tilde{\zeta} + 16\tilde{c}_{5})}{2u_0^3}
    \bar{\psi}_x \sum_{i=1}^{3} \Sigma_{i} B_{x,i} \psi_{x} \CL
  &+\frac{\tilde{c}_{EE}}{2u_0^4}
  \bar{\psi}_x \sum_{i=1}^{3} \gamma_{i} [ (T_{4} - T_{-4}), E_{x,i} ] \psi_{x} \CL
  &+i\frac{\tilde{c}_{3}}{2u_0^4}
  \bar{\psi}_x \sum_{i=1}^{3} \gamma_{i} \Sigma_{i} 
    \{ (T_{i}-T_{-i}), B_{x,i} \} \psi_{x} \CL
  &+i\frac{\tilde{c}_{3}}{2u_0^4} 
  \bar{\psi}_x \sum_{i=1}^{3} \sum_{j \neq i}^{3} \gamma_{i} \Sigma_{j} 
    [ (T_{i}-T_{-i}), B_{x,j} ] \psi_{x} \CL
  &+i\frac{\tilde{c}_{5}}{u_0^4}
  \bar{\psi}_x \sum_{i=1}^{3} 
    \left( -\frac{1}{4} \Sigma_{i} T^{(3)}_{i} \vphantom{\sum_{j \neq i}^{3}} \right. \CL
     &+ \left. \sum_{j \neq i}^{3} 
            \{ \Sigma_{i} B_{x,i}, (T_{j} + T_{-j}) \}
    \right)
    \psi_{x} ,
  \label{eq:OKtad}
\end{align}
where 
\begin{align}
    T^{(3)}_{i} \equiv \sum_{j,k=1}^{3} \epsilon_{ijk}
    \left[ T_{-k}(T_{j}-T_{-j})T_{k} - T_{k}(T_{j}-T_{-j})T_{-k} \right]
\end{align}
is the three-link part of the $c_{5}$ term (which is composed of three- and five-link terms), $i=1,2,3$
denote the spatial directions, and a sum over sites $x\in\mathbb{Z}^4$ is implied.
The matrices $\alpha_i$ and $\Sigma_i$ are defined by
\begin{align}
  \sigma^{\mu\nu} &= \frac{i}{2} [ \gamma^{\mu},\gamma^{\nu} ] , \\
  \sigma^{4i}     &= i\sigma^{0i} = i\alpha^i , \\
  \sigma^{ij}     &= -\epsilon^{ijk}\Sigma_{k} ,
\end{align}
where $\gamma^\mu$ are the Dirac matrices, and the totally anti-symmetric tensor component
$\epsilon^{123}=1$.
We have coded this form of the tadpole-improved OK action with the USQCD~\cite{USQCD:web} software QOPQDP
and the MILC code, with the optimization scheme discussed in Ref.~\cite{Jang:LAT2013}.

The tadpole-improved couplings $\tilde{c}_i$ are obtained by applying the matching conditions in
Ref.~\cite{Oktay2008:PhysRevD.78.014504}, substituting $\tilde{m}_0a$ for $m_0a$, $\tilde{r}_s$ for $r_s$,
$\tilde{\zeta}$ for $\zeta$, and using Eq.~\eqref{eq:kappa}; in practice, $\tilde{r}_s = r_s$ and
$\tilde{\zeta}=\zeta$.
The redundant coefficient of the Wilson term, $r_s$, is set to one to lift the unwanted fermion doublers as
usual.
The tuning parameter $\zeta=\kappa_s/\kappa_t$ plays the following role.
Recall the definitions of the rest mass $m_1$ and the kinetic mass $m_2$, from the energy~$E(\bm{p})$:
\begin{align}
    m_1 &= E(\bm{0}) ,
    \label{eq:rest} \\
    \frac{1}{m_2} &= \left.\frac{\partial^2 E}{\partial p_i^2} \right|_{\bm{p}=\bm{0}} . 
    \label{eq:kin}
\end{align}
At the tree level~\cite{EKM,Oktay2008:PhysRevD.78.014504},
\begin{align} 
    m_1a &= \ln (1 + m_0a) ,
    \label{eq:tune_m1} \\
    \frac{1}{m_2a} &= \frac{ r_{s} \zeta }{ 1+m_0a } + \frac{ 2 \zeta^{2} }{ m_0a(2+m_0a) } ,
    \label{eq:tune_m2}
\end{align}
for both Fermilab and OK actions.
Substituting $\tilde{m}_0a$ from Eq.~(\ref{eq:kappa}) for $m_0a$ in Eqs.~(\ref{eq:tune_m1})
and~(\ref{eq:tune_m2}) yields tadpole-improved tree-level masses, denoted $\tilde{m}_1$ and~$\tilde{m}_2$. 
One can arrange for $m_2=m_1$ (or $\tilde{m}_1=\tilde{m}_2$) by tuning~$\zeta$.
In fact, this tuning can be carried out nonperturbatively with hadron masses---denoted in this paper as
$M_1$ and~$M_2$.
Unfortunately, in that case the discretization errors of $M_2$ are passed on to $M_1$.

In this paper, we set $\zeta=1$, which is just the Fermilab nonrelativistic interpretation~\cite{EKM}, in
which $m_2$ is taken as the physically relevant mass.
In the splittings of hadron rest masses, the quark contribution $m_1$ cancels; the combinations of masses of
the OK action presented in Secs.~\ref{sec:icp} and~\ref{sec:hfs} are based on splittings and, hence, are
immune to this choice.
(The quark rest mass also does not affect combinations of matrix elements of interest
here~\cite{Kronfeld:PhysRevD.62.014505}.)

Note that the rest mass $m_1$ and kinetic mass $m_2$ have the bare-mass dependences in
Eqs.~\eqref{eq:tune_m1} and \eqref{eq:tune_m2} for both the OK and Fermilab actions.
However, the hopping parameter $\kappa$, Eq.~\eqref{eq:kappa}, differs by the $c_4$ term in
Eq.~\eqref{eq:OKtad}, which, along with the $c_1$ and $c_2$ terms, improves the $\order((a\bm{p})^4)$ terms
in the dispersion relation \cite{Oktay2008:PhysRevD.78.014504}.


\section{Meson Correlators}
\label{sec:corr}

\subsection{Data Description}

We use the MILC asqtad-staggered $N_f=2+1$ gauge ensemble that has dimensions $N_L^3\times N_T=20^3\times64$,
$\beta=6.79$, tree-level tadpole factor $u_0=0.8688$ from the plaquette, and lattice spacing
$a\approx0.12$~fm \cite{Bazavov:RevModPhys.82.1349}.
The asqtad-staggered action \cite{Naik1986:NuclPhysB.316.238,Blum1997:PhysRevD.55.1133,%
Orginos1998:PhysRevD.59.014501,Lepage1998:PhysRevD.59.074502,Orginos1999:PhysRevD.60.054503} is used for
the light degenerate sea quarks with mass $am_l=0.02$ and strange sea quark with mass $am_s=0.05$.
For the tests reported here, we use $N_\text{cfg}=500$ of the approximately 2000 available configurations.

On each configuration, we use $N_\text{src}=6$ sources $(t_i,\bm{x}_i)$ for calculating valence quark
propagators.
The source time slices $t_i$ are evenly spaced along the lattice with a randomized offset $t_0\in[0,20)$ for
each configuration.
The spatial source coordinates $\bm{x}_i$ are randomly chosen within the spatial cube.

\begin{table}[tp]
    \caption{Hopping parameters $\kappa_\text{OK}$ for the OK action and $\kappa_\text{FL}$ for the
    Fermilab action.
    The hopping parameters for each action are chosen so that the heavy-light meson masses ($M_{1B_s}$,
    $M_{2B_s}$, $M_{1D_s}$, $M_{2D_s}$) are close to the physical masses $M_{B_s}$ and $M_{D_s}$ 
    converted to lattice units for this ensemble.
    The rest mass $M_1$ and kinetic mass $M_2$ for a meson are defined by the Eqs.~\eqref{eq:rest}
    and~\eqref{eq:kin}, respectively, with meson energies~$E(\bm{p})$.
    The vertically aligned values yield approximately equal heavy-light meson masses.}
  \label{tab:kappa}
  \renewcommand{\arraystretch}{1.1}
    \begin{tabular}{l @{\quad} c c c c @{\quad} c c c c}
      \hline\hline
      \multicolumn{1}{c}{$Q$} &
      \multicolumn{4}{c}{$b$-like} & 
      \multicolumn{4}{c}{$c$-like}
      \\ \hline
      $\kappa_\text{OK}$ & 0.039 & 0.040 & 0.041 & 0.042 & 0.0468 & 0.048 & 0.049 & 0.050 
      \\
      $\kappa_\text{FL}$ & & & 0.083 & 0.091 & & 0.121 & 0.127 &
      \\ \hline\hline
    \end{tabular}
\end{table}

We compute two-point correlators
\begin{equation}
	 C^M(t,\bm{p}) = \sum_{\bm{x}} e^{\mathrm{i}\bm{p} \cdot \bm{x}}
        \expv{\mathcal{O}^{M\dagger}(t,\bm{x}) \mathcal{O}^M(0,\bm{0})} ,
    \label{eq:corr}
\end{equation}
for heavy-light mesons $M=\bar{Q}s$ and quarkonia $M=\bar{Q}Q$ with momentum $\bm{p}$ and heavy
quarks~$Q=b,c$.
The interpolating operators $\mathcal{O}^M(x)$ are
\begin{align}
  \mathcal{O}^{\bar{Q}q}_\mathsf{t}(x) &=
      \bar{\psi}_\alpha(x) \Gamma_{\alpha\beta} \Omega_{\beta\mathsf{t}}(x) \chi(x) ,\\
  \mathcal{O}^{\bar{Q}Q}(x) &= \bar{\psi}_\alpha(x) \Gamma_{\alpha\beta} \psi_{\beta}(x) ,
\end{align}
for heavy-light mesons and quarkonium, respectively.
Here the heavy-quark field $\psi$ is the Wilson-type fermion appearing in the OK or Fermilab action, while
the light-quark field $\chi$ is the staggered fermion appearing in the asqtad-staggered action.
The taste degree of freedom for the staggered fermion is obtained from the one-component field $\chi$ with
$\Omega(x) = \gamma_1^{x_1} \gamma_2^{x_2} \gamma_3^{x_3} \gamma_4^{x_4}$ \cite{Wingate,Bernard}.
We generate pseudoscalar and vector meson correlators with the spin structures $\Gamma = \gamma_5$ and
$\gamma_i$ (for $i=1,2,3$), respectively.

We generate data with ten meson momenta $a\bm{p} = 2\pi\bm{n}/N_L$: %
$\bm{n}=(0,0,0)$, $(1,0,0)$, $(1,1,0)$, $(1,1,1)$, $(2,0,0)$, $(2,1,0)$, $(2,1,1)$, $(2,2,0)$, $(2,2,1)$,
$(3,0,0)$, including all permutations of the components.
These momenta all satisfy $(a\bm{p})^2<0.9$.

The hopping parameter values to simulate $b$-like and $c$-like quarks are given in Table~\ref{tab:kappa}.
We fix the valence light quark mass for the heavy-light meson correlators to the strange sea quark mass
$am_s$.
Hence, we refer to the heavy-light mesons as ``$B_s$'' or ``$D_s$'', depending on the $\kappa$ value for the
heavy quark in Table~\ref{tab:kappa}.
These labels are merely convenient: the tests carried out below do not rely on the physical interpretation
of the mesons.
In anticipation of tuning runs for the OK action, we generate $B_s$ and $D_s$ correlators by using the OK
action with four different values for the hopping parameter $\kappa_\text{OK}$.
For purposes of comparison, we simulate with the Fermilab action with two values for the hopping
parameter~$\kappa_\text{FL}$.

\subsection{Correlator Fits}

\begin{table}[tp]
    \caption{Bayesian priors used for two-point correlator fits.
    For heavy-light mesons, the same priors are used for both heavy quarks $Q=c,b$.
    For bottomonium, we present a fit without the excited state, i.e., $R_1=0$ in Eq.~\eqref{eq:corrfitfn}.}
    \label{tab:2pt-priors}
    \renewcommand{\arraystretch}{1.2}
    \begin{ruledtabular}
    \begin{tabular}{ccccccc}
    Meson   &
    $R_1$   & $\Delta E_1$   &
    $R_0^p$ & $\Delta E^p_0$ &
    $R_1^p$ & $\Delta E^p_1$
    \\ \hline
    $\wbar{Q}q$ & 3.0(1.5) & 0.5(3) & 0.2(2) & 0.1(2) & 3.0(1.5) & 0.5(4) \\
    $\wbar{c}c$ & 1.5(1.0) & 0.5(4) &  &  &  &  \\
  \end{tabular}
  \end{ruledtabular}
\end{table}
The ground state energies are extracted from correlator fits to the function
%
%
\begin{align}
    f(t) &= g(t) + g(T-t) , \label{eq:T-t} \\
    g(t) &= \sum_{i=0,1} \left[A_i e^{-E_i t} - (-1)^t A_i^p e^{-E_i^p t} \right] \nonumber \\
         &= A_0 e^{-E_0t} \sum_{i=0,1} \left[R_i e^{-\Delta E_it} -(-1)^t R_i^p e^{-\Delta E_i^pt} \right],
    \label{eq:corrfitfn} 
\end{align}
where $A_0$ is the ground-state amplitude, and $A_1$ is the first excited-state amplitude.
We also incorporate the staggered parity partner state with amplitude $A_i^p$ and energy $E_i^p$ into the
fit function.
In practice, we take an amplitude ratio $R_i^{(p)}=A^{(p)}_i/A_0$ and energy difference $\Delta
E^{(p)}_i=E^{(p)}_i-E^{(p)}_{i-1}$ as fit parameters instead of $A^{(p)}_i$ and $E^{(p)}_i$.
By definition $R_0=1$, $\Delta E_0=0$, $E^p_{-1} \equiv E_0$, and $R_0$ and $\Delta E_0$ are not fit
parameters.
The parity partners are not present for quarkonium, because they do not contain staggered fermions, so we
set $R_i^p=0$ for quarkonium.
The fits are carried out with the Bayesian priors given in Table~\ref{tab:2pt-priors}.

\begin{figure}[t!]
    \centering
 \vspace{3em}
   \subfigure[~Residual]{%
        \includegraphics[width=0.9\columnwidth]{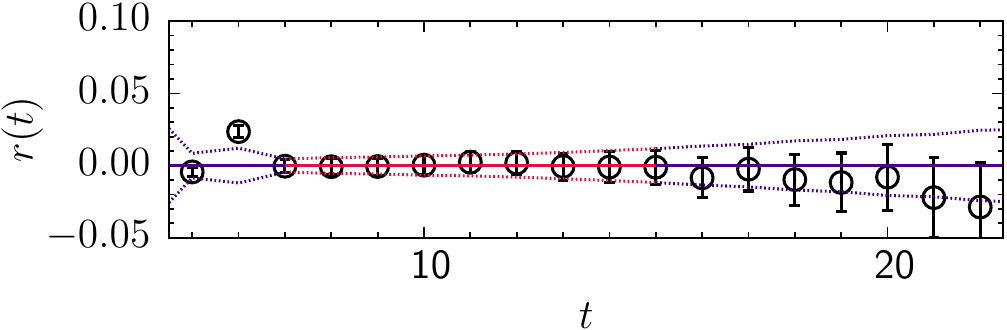}
        \label{fig:hl-resd}
    }
    \subfigure[~Effective Mass]{%
        \includegraphics[width=0.9\columnwidth]{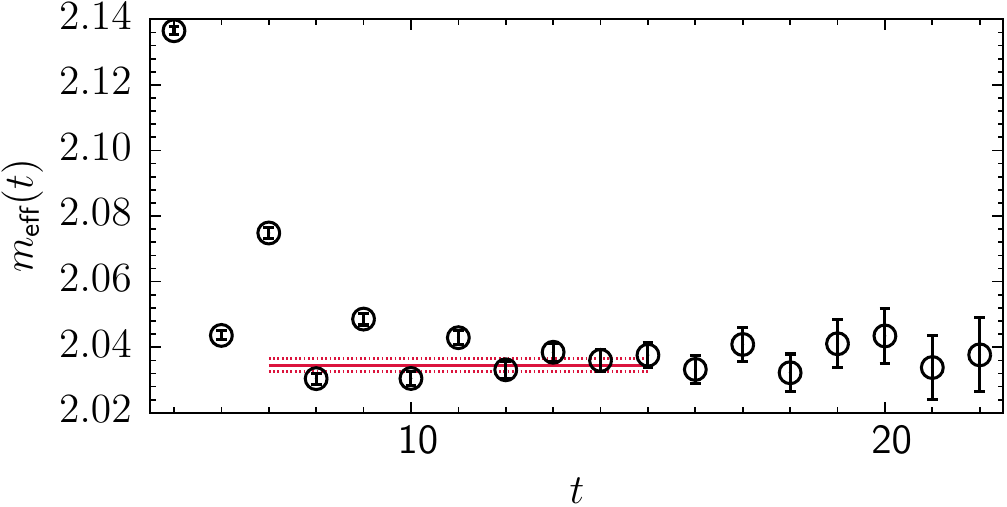}
        \label{fig:hl-meff}
    }
    \caption{Correlated fit results for a pseudoscalar heavy-light meson correlator generated with the OK 
        action at $\kappa=0.041$ and $\bm{p}=\bm{0}$.
        In Fig.~\ref{fig:hl-resd}, the outer dotted lines show the statistical error of the fit.
        In Fig.~\ref{fig:hl-meff}, the fit result for the ground-state energy $E$ is denoted by the 
        horizontal red line plotted over the chosen fit interval.}  
    \label{fig:corr-fit-hl-ps}
\vspace{2em}
    \centering
    \subfigure[~Residual]{%
        \includegraphics[width=0.9\columnwidth]{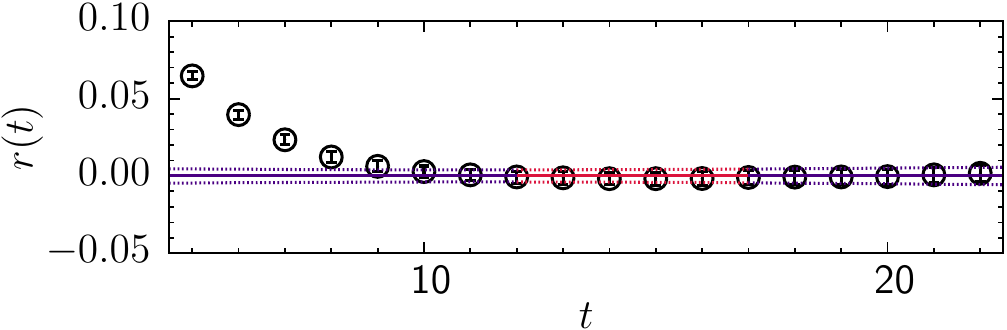}
        \label{fig:hh-resd}
    }
    \subfigure[~Effective Mass]{%
        \includegraphics[width=0.9\columnwidth]{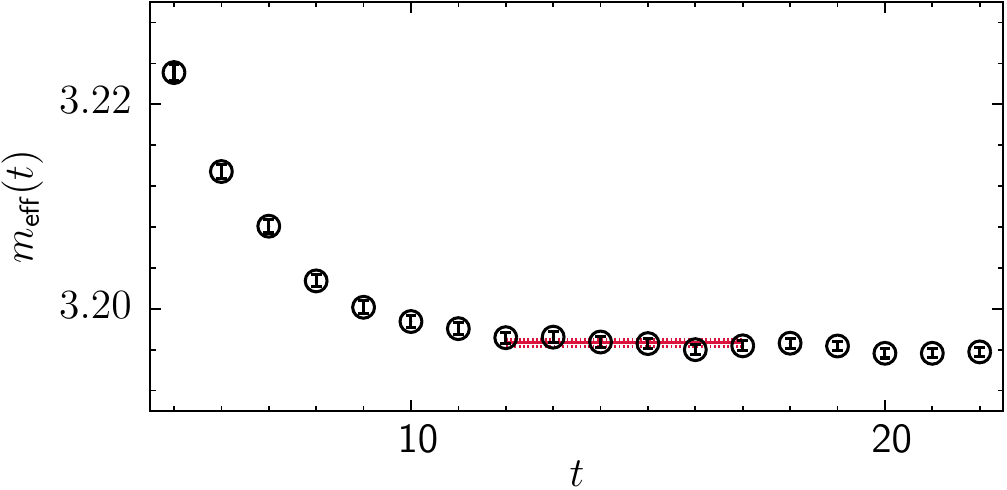}
        \label{fig:hh-meff}
    }
    \caption{Correlated fit results for a pseudoscalar bottomonium correlator generated with the OK action
        at $\kappa=0.041$ and $\bm{p}=\bm{0}$.
        In Fig.~\ref{fig:hh-resd}, the outer dotted lines show the statistical error of the fit.
        In Fig.~\ref{fig:hh-meff}, the fit result for the ground-state energy $E$ is denoted by the 
        horizontal red line plotted over the chosen fit interval.}
    \label{fig:corr-fit-hh-ps}
\end{figure}

\begin{table*}[tp]
    \caption{Dispersion fit results for heavy-light mesons with masses near those of the $D_s^{(\ast)}$ and
    $B_s^{(*)}$.
    The energies $E$ are from the correlator fits to the function in Eq.~\eqref{eq:disp}.
    The first column indicates the lattice action and the (approximate) heavy-quark flavor.
    The second column is the hopping parameter $\kappa$.
    The following columns are the rest mass $M_1$, the kinetic mass $M_2$, the generalized masses $M_4$ and
    $M_6$, and the coefficient of an $O(3)$ symmetry breaking term $W_4$.
    The last two columns are the $\chi^2$ divided by the degrees of freedom (dof)
    $N_\text{data}-N_\text{param}$, $p$-value.
    All ten momenta data points are included in the dispersion relation fit; these spectra are obtained 
    from the correlator fits to Eq.~\eqref{eq:corrfitfn} which include all correlations.
    Fits to the dispersion relation include all correlations among different momenta.  
    Errors are all from a single-elimination jackknife.}
\label{tab:dispfit-hl}
\renewcommand{\arraystretch}{1.1}
\subtable[~Pseudoscalar]{
\label{tab:dispfit-ps-hl}
\begin{ruledtabular}
  \begin{tabular}{c.{2}.{3}.{3}.{3}.{3}.{3}.{4}.{4}}
    \multicolumn{1}{c}{Action ($Q$)}
    & \multicolumn{1}{c}{$\kappa$}
    & \multicolumn{1}{c}{$M_1$}
    & \multicolumn{1}{c}{$M_2$} 
    & \multicolumn{1}{c}{$M_4$} 
    & \multicolumn{1}{c}{$W_4 \times 10^2$}
    & \multicolumn{1}{c}{$M_6$}
    & \multicolumn{1}{c}{$\chi^2/\text{dof}$} 
    & \multicolumn{1}{c}{$p$}
    \\ \hline
    \multirow{4}{*}{OK ($b$-like)} 
& 0.039 & 2.2141(24) & 3.924(63) & 2.50(16) & -6.1(7)   & \multicolumn{1}{c}{--} & 1.13(4) & 0.34(2)  \\ 
& 0.040 & 2.1261(22) & 3.580(53) & 2.46(15) & -5.5(8)   & \multicolumn{1}{c}{--} & 1.21(4) & 0.30(2)  \\ 
& 0.041 & 2.0345(19) & 3.256(45) & 2.40(15) & -4.9(9)   & \multicolumn{1}{c}{--} & 1.26(4) & 0.27(2)  \\ 
& 0.042 & 1.9382(17) & 2.952(37) & 2.33(15) & -4.0(1.0) & \multicolumn{1}{c}{--} & 1.35(4) & 0.23(2)  \\ 
    \hline
    \multirow{4}{*}{OK ($c$-like)} 
& 0.0468 & 1.3742(9) & 1.663(12) & 1.59(7) &  3.0(2.4) & \multicolumn{1}{c}{--} & 1.29(4) & 0.26(2)  \\ 
& 0.048 & 1.1861(8) & 1.362(8)  & 1.33(4) &  6.8(3.5) & 1.4(2) & 1.47(5) & 0.20(2) \\
& 0.049 & 1.0003(7) & 1.104(6)  & 1.12(3) &  7.3(5.5) & 1.2(1) & 1.47(5) & 0.20(2)  \\ 
& 0.050 & 0.7688(5) & 0.827(5)  & 0.85(2) &  7.4(10.2) & 1.0(1) & 0.32(2) & 0.90(1) \\ 
    \hline
    \multirow{2}{*}{FL ($b$-like)}
& 0.083 & 2.0668(31) & 3.250(83) & 2.47(46) & 2.4(1.3) & \multicolumn{1}{c}{--} & 0.50(3) & 0.81(2)  \\ 
& 0.091 & 1.8835(28) & 2.746(70) & 2.28(43) & 3.3(1.8) & \multicolumn{1}{c}{--} & 0.48(3) & 0.83(2)  \\ 
    \hline
    \multirow{2}{*}{FL ($c$-like)}
    & 0.121 & 1.1533(8) & 1.308(9) & 1.20(4) & 12.9(4.5) & 1.4(2) & 1.38(5) & 0.23(2) \\ 
& 0.127 & 0.9736(6) & 1.048(9) & 0.97(5) & 16.9(7.2) & 1.1(1) & 0.83(4) & 0.53(3)  \\ 
  \end{tabular}
  \end{ruledtabular}
  }
  \renewcommand{\arraystretch}{1.1}
  \subtable[~Vector]{
  \label{tab:dispfit-v-hl}
  \begin{ruledtabular}
  \begin{tabular}{c.{2}.{3}.{3}.{3}.{3}.{3}.{4}.{4}}
    \multicolumn{1}{c}{Action ($Q$)}
    & \multicolumn{1}{c}{$\kappa$}
    & \multicolumn{1}{c}{$M_1$}
    & \multicolumn{1}{c}{$M_2$} 
    & \multicolumn{1}{c}{$M_4$} 
    & \multicolumn{1}{c}{$W_4 \times 10^2$}
    & \multicolumn{1}{c}{$M_6$}
    & \multicolumn{1}{c}{$\chi^2/\text{dof}$} 
    & \multicolumn{1}{c}{$p$}
    \\ \hline
    \multirow{4}{*}{OK ($b$-like)} 
& 0.039 & 2.2330(31) & 3.988(87) & 2.69(30) & -4.5(1.0) & \multicolumn{1}{c}{--} & 1.36(4) & 0.23(2)  \\ 
& 0.040 & 2.1475(28) & 3.653(74) & 2.60(28) & -4.1(1.1) & \multicolumn{1}{c}{--} & 1.23(4) & 0.29(2)  \\ 
& 0.041 & 2.0589(25) & 3.325(61) & 2.53(27) & -3.7(1.2) & \multicolumn{1}{c}{--} & 1.20(4) & 0.31(2)  \\ 
& 0.042 & 1.9662(23) & 3.016(50) & 2.47(27) & -3.1(1.3) & \multicolumn{1}{c}{--} & 1.18(4) & 0.31(2)  \\ 
    \hline
    \multirow{4}{*}{OK ($c$-like)} 
& 0.0468 & 1.4324(14) & 1.733(18) & 1.74(14) & 2.9(3.6)  & \multicolumn{1}{c}{--} & 1.04(4) & 0.39(2)  \\ 
& 0.048 & 1.2594(13) & 1.443(15) & 1.48(10) & 4.5(5.7)  & \multicolumn{1}{c}{--} & 0.97(4) & 0.45(2)  \\ 
& 0.049 & 1.0921(12) & 1.205(13) & 1.27(9) &  4.2(9.6)  & \multicolumn{1}{c}{--} & 0.78(3) & 0.59(3)  \\ 
& 0.050 & 0.8920(14) & 0.969(15) & 0.98(7) & -8.8(21.7) & \multicolumn{1}{c}{--} & 0.35(2) & 0.91(1)  \\ 
    \hline
    \multirow{2}{*}{FL ($b$-like)}
& 0.083 & 2.0906(53) & 3.406(132) & 3.18(2.29) & 0.6(1.9) & \multicolumn{1}{c}{--} & 0.27(2) & 0.95(1)  \\ 
& 0.091 & 1.9157(37) & 2.890(74)  & 2.68(91)   & 1.0(2.9) & \multicolumn{1}{c}{--} & 0.54(3) & 0.78(2)  \\ 
    \hline
    \multirow{2}{*}{FL ($c$-like)}
& 0.121 & 1.2287(15) & 1.438(19) & 1.43(10) & 10.7(5.5) & \multicolumn{1}{c}{--} & 0.58(3) & 0.75(2)  \\ 
& 0.127 & 1.0693(16) & 1.206(17) & 1.24(8)  & 12.7(9.2) & \multicolumn{1}{c}{--} & 0.67(3) & 0.67(2)  \\ 
  \end{tabular}
  \end{ruledtabular}
  }
\end{table*}

\begin{table*}[tp]
    \caption{Dispersion fit results for quarkonium.
    The energies $E$ are from the correlator fits to the function in Eq.~\eqref{eq:disp}.
    The first column indicates the lattice action and the (approximate) heavy-quark flavor.
    The second column is the hopping parameter $\kappa$.
    The following columns are the rest mass $M_1$, the kinetic mass $M_2$, the generalized masses $M_4$ and
    $M_6$, and the coefficient of an $O(3)$ symmetry breaking terms $W_4$ and $W_6^\prime$.
    The last two columns are the $\chi^2$ divided by the degrees of freedom (dof)
    $N_\text{data}-N_\text{param}$, $p$-value.
    All ten momenta data points are included in the dispersion relation fit; these spectra are obtained from the
    correlator fits to Eq.~\eqref{eq:corrfitfn}, setting $R_{i}^{p} =0$, which include all correlations; $R_1=0$
    for bottomonium fits.
    Fits to the dispersion relation include all correlations among different momenta.
    Errors are all from a single-elimination jackknife.}
    \label{tab:dispfit-hh}
    \renewcommand{\arraystretch}{1.1}
    \subtable[~Pseudoscalar]{
  \label{tab:dispfit-ps-hh}
  \begin{ruledtabular}
  \begin{tabular}{c.{3}.{5}.{5}.{3}.{5}.{2}.{4}.{5}.{5}}
    \multicolumn{1}{c}{Action ($Q$)}
    & \multicolumn{1}{c}{$\kappa$}
    & \multicolumn{1}{c}{$M_1$}
    & \multicolumn{1}{c}{$M_2$} 
    & \multicolumn{1}{c}{$M_4$} 
    & \multicolumn{1}{c}{$W_4 \times 10^2$}
    & \multicolumn{1}{c}{$M_6$}
    & \multicolumn{1}{c}{$W_6^\prime \times 10^4$}
    & \multicolumn{1}{c}{$\chi^2/\text{dof}$} 
    & \multicolumn{1}{c}{$p$}
    \\ \hline
    \multirow{4}{*}{OK ($b$-like)} 
& 0.039 & 3.5200(4) & 6.637(19) & 4.58(8) & -0.91(3) & \multicolumn{1}{c}{--} & \multicolumn{1}{c}{--} & 0.51(3) & 0.80(2)  \\ 
& 0.040 & 3.3616(4) & 6.120(17) & 4.38(8) & -0.97(3) & \multicolumn{1}{c}{--} & \multicolumn{1}{c}{--} & 0.48(3) & 0.82(2)  \\ 
& 0.041 & 3.1968(4) & 5.601(16) & 4.16(8) & -0.99(4) & \multicolumn{1}{c}{--} & \multicolumn{1}{c}{--} & 0.95(4) & 0.46(3)  \\ 
& 0.042 & 3.0242(4) & 5.083(14) & 3.90(7) & -0.97(5) & \multicolumn{1}{c}{--} & \multicolumn{1}{c}{--} & 1.32(4) & 0.25(2)  \\ 
    \hline
    \multirow{4}{*}{OK ($c$-like)} 
& 0.0468 & 2.0169(5) & 2.647(11) & 2.41(9) &  0.7(4)   & \multicolumn{1}{c}{--} & \multicolumn{1}{c}{--} & 0.59(3) & 0.74(2)  \\ 
& 0.048 & 1.6819(5) & 2.063(8)  & 1.97(6) &  2.1(7)   & \multicolumn{1}{c}{--} & \multicolumn{1}{c}{--} & 0.82(3) & 0.55(3)  \\ 
& 0.049 & 1.3516(5) & 1.574(5)  & 1.54(3) &  3.6(1.5) & \multicolumn{1}{c}{--} & \multicolumn{1}{c}{--} & 1.20(4) & 0.30(2)  \\ 
& 0.050 & 0.9419(5) & 1.056(4)  & 1.08(2) & -0.6(4.1) & \multicolumn{1}{c}{--} & \multicolumn{1}{c}{--} & 1.36(4) & 0.23(2)  \\ 
    \hline
    \multirow{2}{*}{FL ($b$-like)}
& 0.083 & 3.1765(3) & 9.392(23) & 3.69(2) & 1.84(2)   & 4.1(1) & 5.6(3) & 0.77(3) & 0.55(2)  \\ 
& 0.091 & 2.8764(3) & 6.894(18) & 3.19(2) & 2.45(4)   & 4.0(3) & 9.2(6) & 0.75(4) & 0.56(3)  \\ 
    \hline
    \multirow{2}{*}{FL ($c$-like)}
& 0.121 & 1.6216(5) & 2.128(9) & 1.69(3)   & 8.5(7)    & 1.8(1)  & \multicolumn{1}{c}{--} & 1.39(5) & 0.23(2)  \\ 
& 0.127 & 1.3087(5) & 1.561(6) & 1.35(2)   & 12.0(1.5) & 1.4(1)  & \multicolumn{1}{c}{--} & 0.83(4) & 0.53(3)  \\ 
  \end{tabular}
  \end{ruledtabular}
  }
  \renewcommand{\arraystretch}{1.1}
  \subtable[~Vector]{
  \label{tab:dispfit-v-hh}
  \begin{ruledtabular}
  \begin{tabular}{c.{3}.{5}.{5}.{3}.{5}.{2}.{4}.{5}.{5}}
    \multicolumn{1}{c}{Action ($Q$)}
    & \multicolumn{1}{c}{$\kappa$}
    & \multicolumn{1}{c}{$M_1$}
    & \multicolumn{1}{c}{$M_2$} 
    & \multicolumn{1}{c}{$M_4$} 
    & \multicolumn{1}{c}{$W_4 \times 10^2$}
    & \multicolumn{1}{c}{$M_6$}
    & \multicolumn{1}{c}{$W_6^\prime \times 10^4$}
    & \multicolumn{1}{c}{$\chi^2/\text{dof}$} 
    & \multicolumn{1}{c}{$p$}
    \\ \hline
    \multirow{4}{*}{OK ($b$-like)} 
    & 0.039 & 3.5393(4) & 6.626(20) & 4.69(9) & -0.99(3) & \multicolumn{1}{c}{--} & \multicolumn{1}{c}{--} & 0.64(3) & 0.70(2)  \\ 
    & 0.040 & 3.3827(4) & 6.116(19) & 4.48(10) & -1.04(4) & \multicolumn{1}{c}{--} & \multicolumn{1}{c}{--} & 0.69(3) & 0.66(2)  \\ 
    & 0.041 & 3.2201(4) & 5.606(17) & 4.25(10) & -1.05(5) & \multicolumn{1}{c}{--} & \multicolumn{1}{c}{--} & 1.03(4) & 0.40(2)  \\ 
    & 0.042 & 3.0500(4) & 5.096(16) & 3.98(9) & -1.00(6) & \multicolumn{1}{c}{--} & \multicolumn{1}{c}{--} & 1.36(4) & 0.23(2)  \\ 
    \hline
    \multirow{4}{*}{OK ($c$-like)} 
    & 0.0468 & 2.0657(7) & 2.686(14) & 2.55(12) &  1.1(5)   & \multicolumn{1}{c}{--} & \multicolumn{1}{c}{--} & 0.43(2) & 0.86(2)  \\ 
    & 0.048 & 1.7425(7) & 2.112(10) & 2.08(9)  &  2.8(9)   & \multicolumn{1}{c}{--} & \multicolumn{1}{c}{--} & 0.49(3) & 0.81(2)  \\ 
    & 0.049 & 1.4280(7) & 1.640(8)  & 1.63(5)  &  4.5(1.7) & \multicolumn{1}{c}{--} & \multicolumn{1}{c}{--} & 0.66(3) & 0.68(2)  \\ 
    & 0.050 & 1.0497(9) & 1.154(7)  & 1.15(3)  & -1.7(5.9) & \multicolumn{1}{c}{--} & \multicolumn{1}{c}{--} & 0.76(3) & 0.60(3)  \\ 
    \hline
    \multirow{2}{*}{FL ($b$-like)}
& 0.083 & 3.1939(3) & 9.589(25) & 3.70(2) & 1.82(2)  & 4.0(1) & 5.4(3) & 1.53(6) & 0.19(2)  \\ 
& 0.091 & 2.8987(4) & 7.035(20) & 3.21(2) & 2.43(4)  & 4.0(3) & 9.0(5) & 0.44(3) & 0.78(2)  \\ 
    \hline
    \multirow{2}{*}{FL ($c$-like)}
    & 0.121 & 1.6774(8) & 2.235(13) & 1.75(5) & 7.6(9)    & 1.9(2) & \multicolumn{1}{c}{--} & 1.05(4) & 0.39(2)  \\ 
    & 0.127 & 1.3808(9) & 1.680(11) & 1.43(4) & 10.0(2.0) & 1.6(1) & \multicolumn{1}{c}{--} & 0.50(3) & 0.78(2)  \\ 
  \end{tabular}
  \end{ruledtabular}
  }
\end{table*}

We carry out correlated fits with a statistical estimate of the covariance matrix
$\mathrm{cov}(t,t^\prime)$ between different time slices $t$ and $t^\prime$,
\begin{equation}
    \mathrm{cov}(t,t^\prime) = \frac{1}{\mathcal{N}} \sum_{i=1}^{N_\text{cfg}}
        \left[C_i(t) - C(t)\right]  \left[C_i(t^\prime) - C(t^\prime)\right] ,
\end{equation}
where the normalization factor $\mathcal{N}=N_\text{cfg}(N_\text{cfg}-1)$, $C_i(t)$ represents the average
over the $N_\text{src}$ sources for the $i^\text{th}$ gauge field, and the two-point correlator
$C(t)=C^M(t,\bm{p})$ for a given meson type~$M$ and momentum~$\bm{p}$ is estimated with the ensemble average
\begin{equation}
    C(t) = \frac{1}{N_\text{cfg}} \sum_{i=1}^{N_\text{cfg}} C_i(t) .
\end{equation}

To increase statistics, we take advantage of the structure of Eq.~(\ref{eq:T-t}) and average the data for
time separations $t$ and $T-t$.
Then, we take the fit interval $[t_\text{min},t_\text{max}]$, where $0<t_\text{min}<t_\text{max}<T/2$, equal
to $[7,15]$ for the heavy-light meson with the OK action and $[9,15]$ with the Fermilab action.
For charmonium, the fit interval is $[9,14]$ for both actions.
For bottomonium, the fit interval is $[12,17]$ for the OK action and $[14,19]$ for the Fermilab action.
We fix these intervals for fits to all correlators, independent of hopping parameter $\kappa$,
momentum~$\bm{p}$, and pseudoscalar vs.\ vector channel.

The $t_\text{min}$ are chosen by observing the effective mass $m_\text{eff}(t)$ calculated from the
definition
\begin{equation}
	m_\text{eff}(t) = \frac{1}{2} \ln \left(\frac{C^M(t)}{C^M(t+2)}\right)
\end{equation}
as well as comparing the fit results $E$ with the $m_\text{eff}(t)$.
Figure~\ref{fig:corr-fit-hl-ps} shows the effective mass $m_\text{eff}(t)$ and correlator fit residual
$r(t)$
\begin{equation}
	r(t) = \frac{C^M(t)-f(t)}{\abs{C^M(t)}}
\end{equation}
for a pseudoscalar heavy-light meson correlator generated with the OK action, $\kappa_\text{OK}=0.041$, and
$\bm{p}=\bm{0}$.
The pseudoscalar bottomonium correlator fit results obtained with the same action and parameters are given
in Fig.~\ref{fig:corr-fit-hh-ps}.
To estimate the statistical errors, we use a single-elimination jackknife.

\section{Meson Masses}
\label{sec:disp}

The dispersion relation of the mesons has the same form as that of the quarks with the quark masses $m_1$
and $m_2$ given in Eqs.~\eqref{eq:rest} and \eqref{eq:kin}, respectively, replaced by meson masses $M_1$ and
$M_2$.
The mismatch between the meson rest and kinetic masses can be exploited to test the improvement of
nonrelativistically interpreted actions.

We fit the ground-state energy $E(\bm{p})$ in Eq.~\eqref{eq:corrfitfn} for each momentum $\bm{p}$ to
the nonrelativistic dispersion relation to obtain the rest mass $M_1$ and kinetic mass $M_2$ of each
meson.
Including terms up to $\order\!\left(\bm{p}^6\right)$, the dispersion relation is
\begin{align}
    E(\bm{p}) &= M_1 + \frac{\bm{p}^2}{2M_2} - \frac{(\bm{p}^2)^2}{8M_4^3}
    + E_4^\prime + E_6 + E_6^\prime , \label{eq:disp} \\
    E_{4}^\prime &= - \frac{a^3W_4}{6} \sum_i p_i^4 , \label{eq:disp-hoc} \\
    E_{6} &= \frac{(\bm{p}^2)^3}{16M_6^5} , \\
    E_{6}^\prime &= \frac{a^5W_6^\prime}{2} \bm{p}^2 \sum_i p_i^4 - \frac{a^5W_6}{3} \sum_i p_i^6 .
\end{align}
The $M_{4,6}$ are generalized masses; the rest mass $M_1$ and these generalized masses $M_{4,6}$ are
expected~\cite{EKM,Oktay2008:PhysRevD.78.014504} to approach the kinetic mass $M_2$ in the continuum limit.
The $O(3)$ rotation-symmetry-breaking terms are $E_4^\prime$ and $E_6^\prime$.
In the continuum limit, the coefficients $a^3W_4$, $a^5W'_6$, and $a^5W_6$ vanish.
We fit the simulation data for $E(\bm{p})$ to the the right-hand side of Eq.~\eqref{eq:disp}, taking the
full covariance matrix among the different momentum channels, and investigate variations by excluding some
or all of the higher-order terms $E_4^\prime$, $E_6$, and $E_6^\prime$. 
We do not introduce priors here.
The seven fit parameters are $M_1$, $M_2^{-1}$, $M_4^{-3}$, $M_6^{-5}$, $W_4$, $W'_6$, and $W_6$.
We find that we do not need all seven terms in the dispersion relation.
In many cases, it suffices to keep only the first five, while some fits give better $p$ values with the
first six.

For the pseudoscalar heavy-light meson, as the spectrum becomes more relativistic, $M_2$ approaches $M_1$,
and including the $E_6$ term results in a better fit.
Because the vector heavy-light meson spectrum has a larger statistical error than the pseudoscalar meson
spectrum, the $E_6$ term is not only determined to be statistically zero, but also does not improve 
the fit.
The $E_6$ term also improves the fit for the charmonium spectrum with the Fermilab action.

\begin{figure}[bp]
    \centering
    \subfigure[~OK]{%
        \includegraphics[width=0.9\columnwidth]{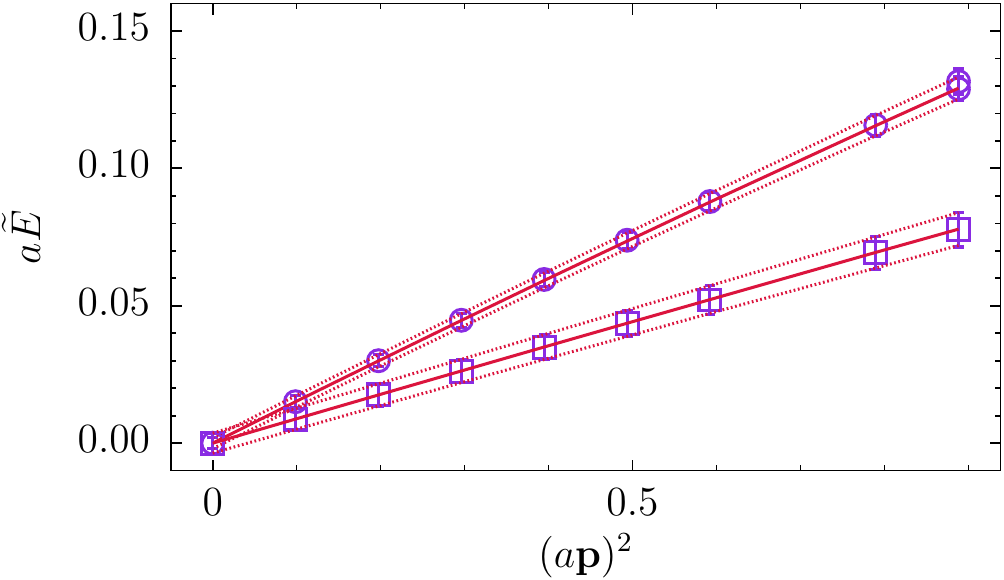}
        \label{fig:disp_ifla_PS}
    }
    \vspace{1em}
    \subfigure[~Fermilab]{%
        \includegraphics[width=0.9\columnwidth]{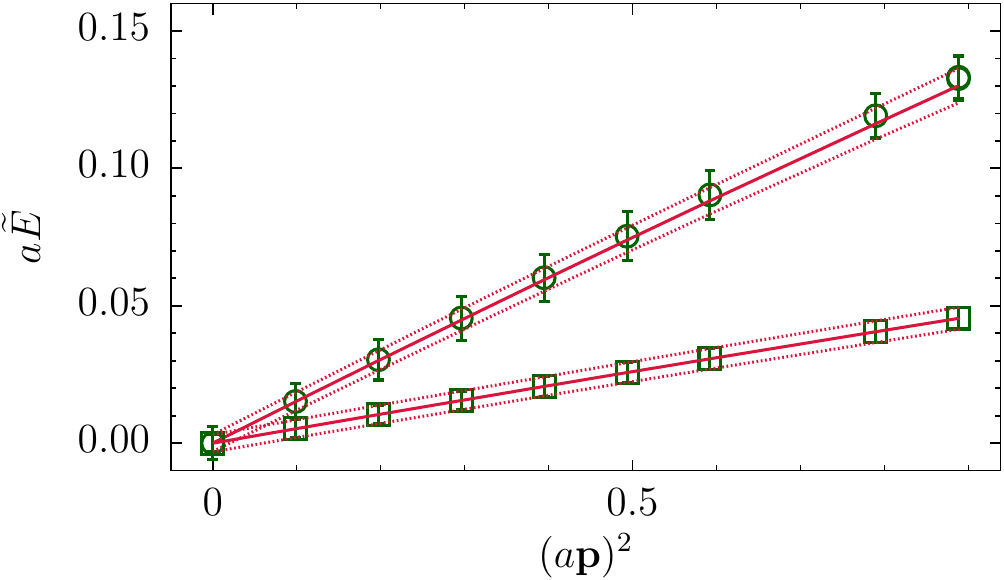}
        \label{fig:disp_clov_PS}
    }
    \caption{The pseudoscalar heavy-light meson (circle) and quarkonium (square) subtracted 
    energies~$\widetilde{E}$ [Eq.~\eqref{eq:Etilde}] as a function of $(a\bm{p})^2$ for %
    (a)~the OK action at $\kappa=0.041$ and %
    (b)~the Fermilab (clover) action at $\kappa=0.083$.
    The lines show fits to Eq.~\eqref{eq:disp}.
    The errors are from a single-elimination jackknife.
    In these plots, the errors for the bottomonium data points and fit lines are scaled by a factor 10.
    The behavior for vector mesons is similar, apart from the larger statistical errors.}
    \label{fig:disp-ps}
\end{figure}

The $W_6^\prime$ term in addition to the $E_6$ term improves the fit for the bottomonium spectrum with the
Fermilab action.
Note that the correlator fit for bottomonium does not include any excited states., although including a
single excited state is statistically consistent, when fitting Eq.~(\ref{eq:disp}) without the $E_6$ and
$E_6^\prime$ terms.
We use the fits with excited states to cross-check the single-state fits.
We observed the same behavior with the two-state fit for the bottomonium spectrum with the OK action.
Note that the bottomonium spectrum for the OK action still results in a good fit without the $E_6$ and
$E_6^{\prime}$ terms, when the correlator fit only accounts for the ground state.

Figure~\ref{fig:disp-ps} shows the dispersion relation fits to the pseudoscalar heavy-light meson and
quarkonium data generated with the OK (Fermilab) action with $\kappa_\text{OK}=0.041$ ($\kappa_\text{FL}=0.083$).
We plot the dispersion relation fit results after subtracting the rest mass and the higher-order
term~$E_4^\prime$.
We thus define
\begin{equation}
    \widetilde{E} = E - M_1 - E_4^\prime
    \label{eq:Etilde}
\end{equation}
to draw the plot.
As one can see from Fig.~\ref{fig:disp-ps}, the slope---and, hence, the kinetic mass~$M_2$---is reliably
determined from fits to Eq.~\eqref{eq:disp}.
Interestingly, the OK action leads to statistical errors noticeably smaller than those from the
Fermilab action, especially for the heavy-light masses.

The fit results for the heavy-light mesons are summarized in Tables~\ref{tab:dispfit-ps-hl}
and~\ref{tab:dispfit-v-hl}.
They are obtained from correlated fits.
From these tables for the heavy-light mesons, one can see that the higher-order generalized mass~$M_4$
approaches the kinetic mass $M_2$ as the kinetic mass decreases.
The rest mass $M_1$ also approaches the kinetic mass $M_2$.
Note that $M_4$ does not necessarily agree better with $M_2$ for the OK action, even though the OK action
tunes the action such that $m_4=m_2$.
A possible explanation is that the binding energy $M_4-m_4$ stems from higher-dimension effects not yet
incorporated into the OK action.

The correlated fit results for quarkonia are summarized in Tables~\ref{tab:dispfit-ps-hh}
and~\ref{tab:dispfit-v-hh}.
For the charmonium results obtained with the OK action, $M_4$ is closer to the kinetic mass $M_2$ than to
the rest mass $M_1$, but for the Fermilab action $M_4$ is closer to $M_1$.
In the bottomonium region, $M_4$ is closer to the rest mass $M_1$ than the kinetic mass $M_2$ for
both actions.
The difference $M_2-M_1$ is large for the Fermilab action at $\kappa_\text{FL}=0.083$, $0.091$, but it is
diminished with the OK action at the comparable values $\kappa_\text{OK}=0.039$, $0.040$, $0.041$, $0.042$.
However, even with the OK action, the difference between $M_4$ and $M_2$ in the bottomonium spectra is
larger than that in the heavy-light meson spectra.

\section{The ``Inconsistency''~\cite{Collins}}
\label{sec:icp}

To study how the Fermilab method works, Ref.~\cite{Collins} introduced the quantity
\begin{equation}
    I = \frac{2{\delta}M_{\wbar{Q}q} - ({\delta}M_{\wbar{Q}Q} + {\delta}M_{\wbar{q}q})}{2M_{2\wbar{Q}q}} ,
    \label{eq:iparam}
\end{equation}
where
\begin{equation}
    \delta M_{\wbar{Q}q} = M_{2\wbar{Q}q} - M_{1\wbar{Q}q}.
    \label{eq:deltaM}
\end{equation}
The combination~$I$ should vanish, even when the quark rest masses $m_1$ are mistuned.%
\footnote{As exhibited in Eq.~(\ref{eq:M1M2}), $M_2$ contains a binding energy, which is sensitive to the 
higher-dimension discretization errors addressed by the OK action.
With the Fermilab action (as sometimes implemented~\cite{{Aoki:2001ra,Christ2006:PhysRevD.76.074505}}),
a hadron-mass-based tuning of $\zeta$ transmits these errors from $M_2$ to~$M_1$.} %
If $I$ does not vanish, it means that the action contains nontrivial discretization effects at higher
order in the HQET or NRQCD power counting, so $I$ is often called the ``inconsistency.'' %
In fact, as we reproduce below, $I\ne 0$ for the Fermilab action~\cite{Collins} (unless $m_2a\ll1$),
as one can see in Eq.~\eqref{eq:NRdeltaB}.

The explanation is as follows~\cite{Kronfeld}.
The meson masses $M_i$ $(i=1,2)$ can be written as a sum of the quark masses $m_i$ and the binding energy:
\begin{equation}
    M_{i\wbar{Q}q} = m_{i\wbar{Q}} + m_{iq} + B_{i\wbar{Q}q} ,
    \label{eq:M1M2}
\end{equation}
which defines $B_i$.
Upon substituting Eq.~(\ref{eq:M1M2}) into Eqs.~\eqref{eq:deltaM} and~\eqref{eq:iparam}, the quark masses
cancel out, and the inconsistency becomes a relation among the binding energies,
\begin{equation}
    I = \frac{2{\delta}B_{\wbar{Q}q} - ({\delta}B_{\wbar{Q}Q} + {\delta}B_{\wbar{q}q})}{2M_{2\wbar{Q}q}},
    \label{eq:iparam2}
\end{equation}
where
\begin{equation}
    \delta B_{\wbar{Q}q} = B_{2\wbar{Q}q} - B_{1\wbar{Q}q}.
    \label{eq:deltaB}
\end{equation}
The quantities in Eqs.~\eqref{eq:deltaM}, \eqref{eq:M1M2}, and \eqref{eq:deltaB} for heavy ($\wbar{Q}Q$) and
light ($\wbar{q}q$) quarkonium are defined similarly.
Because light quarks always have $ma\ll1$, the $\order((ma)^2)$ distinction between rest and kinetic masses
is negligible for them, so we omit $\delta M_{\bar{q}q}$ (or~$\delta B_{\bar{q}q}$) when computing~$I$.
The rest-mass binding energy $B_1$ is sensitive to effects of $\order(\lambda)$ or $\order(v^2)$, while the
kinetic-mass binding energy $B_2$ is sensitive to effects of $\order(\lambda^3)$ or $\order(v^4)$, with
their larger relative discretization errors (for the clover and Fermilab actions).
The discretization errors can be studied in the nonrelativistic limit via the Breit
equation, yielding~\cite{Kronfeld,Bernard}
\begin{align}
    \delta B_{\wbar{Q}q} &= \frac{5}{3} \frac{\expv{\bm{p}^2}}{2\mu_2} \left[ \mu_2 
        \left(\frac{m_{2\wbar{Q}}^2}{m_{4\wbar{Q}}^3} + \frac{m_{2q}^2}{m_{4q}^3} \right) - 1 \right] 
        \label{eq:NRdeltaB} \\
                         & \hspace{1pc} + \frac{4}{3} a^3 \frac{\expv{\bm{p}^2}}{2\mu_2} 
        \mu_2 \left[w_{4\wbar{Q}} m_{2\wbar{Q}}^2 + w_{4q} m_{2q}^2 \right] + \order(\bm{p}^4) 
        \nonumber
\end{align}
in the $S$ wave, where the reduced mass $\mu_2^{-1} = m_{2\wbar{Q}}^{-1} + m_{2q}^{-1}$, and $m_2$, $m_4$,
and $w_4$ are defined through the quark analog of Eq.~\eqref{eq:disp}.
Here, $\bm{p}$ is the relative momentum of $\wbar{Q}$ and $q$ in their center of mass.
The OK action matches $m_4=m_2$ and $w_4=0$ at the tree level, so the two expressions in square brackets are
of order $\alpha_s$.
Because we use a tadpole-improved action, these one-loop errors are expected to be small.

\begin{figure}[tbp]
    \centering
    \subfigure[~Spin average]{
        \label{fig:iparam-avg}
        \includegraphics[width=0.9\columnwidth]{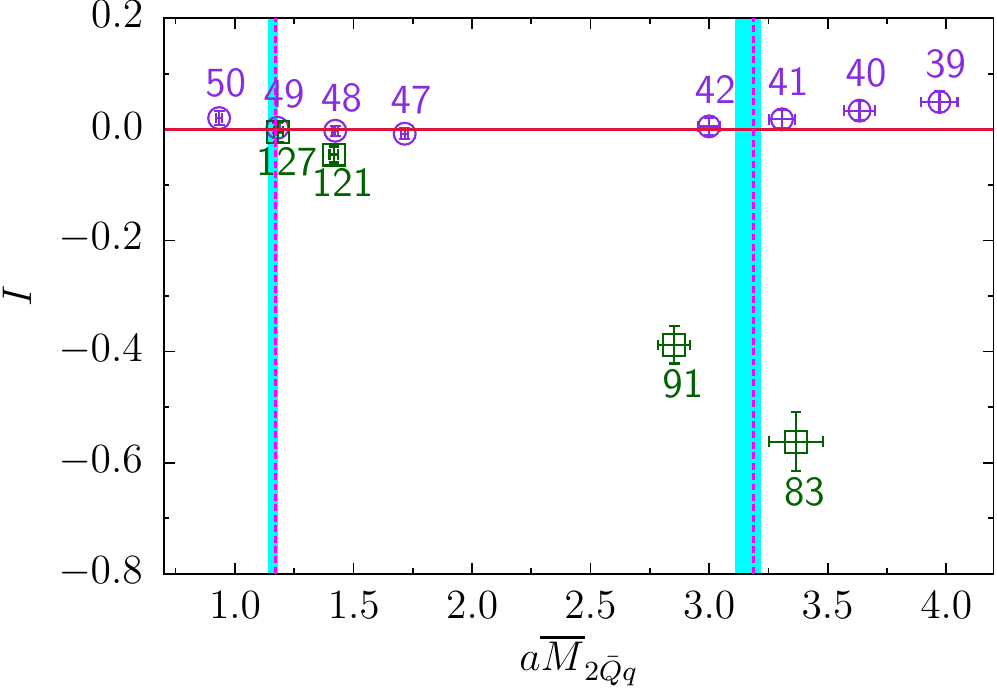}
    }\\
    \vspace*{2em}
    \subfigure[~Pseudoscalar]{
        \label{fig:iparam-ps}
        \includegraphics[width=0.9\columnwidth]{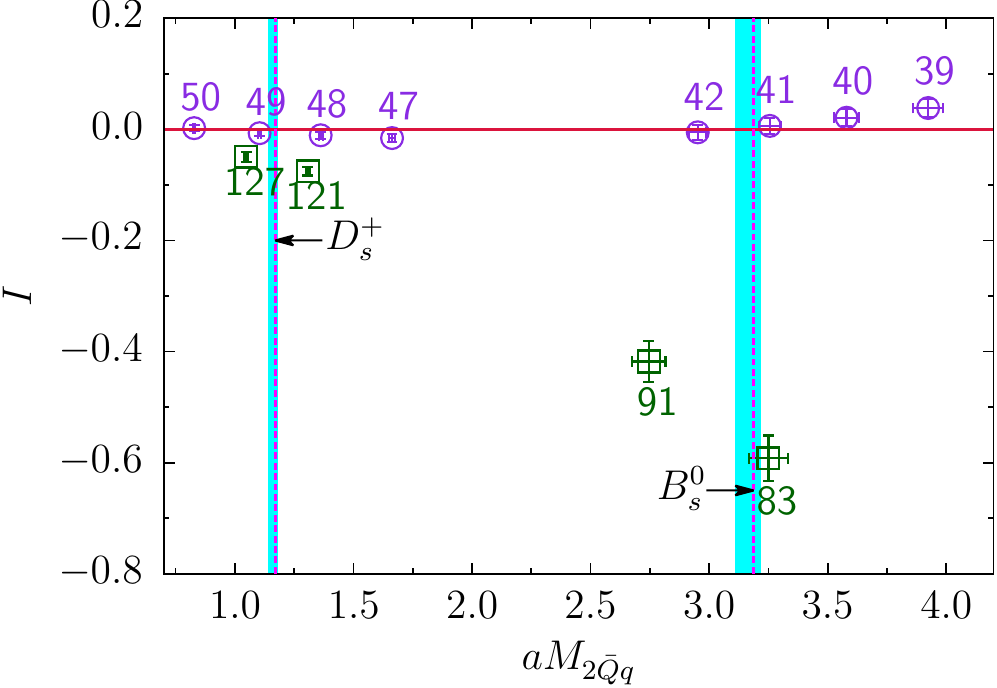}
    }\\
    \vspace*{2em}
    \subfigure[~Vector]{
        \label{fig:iparam-v}
        \includegraphics[width=0.9\columnwidth]{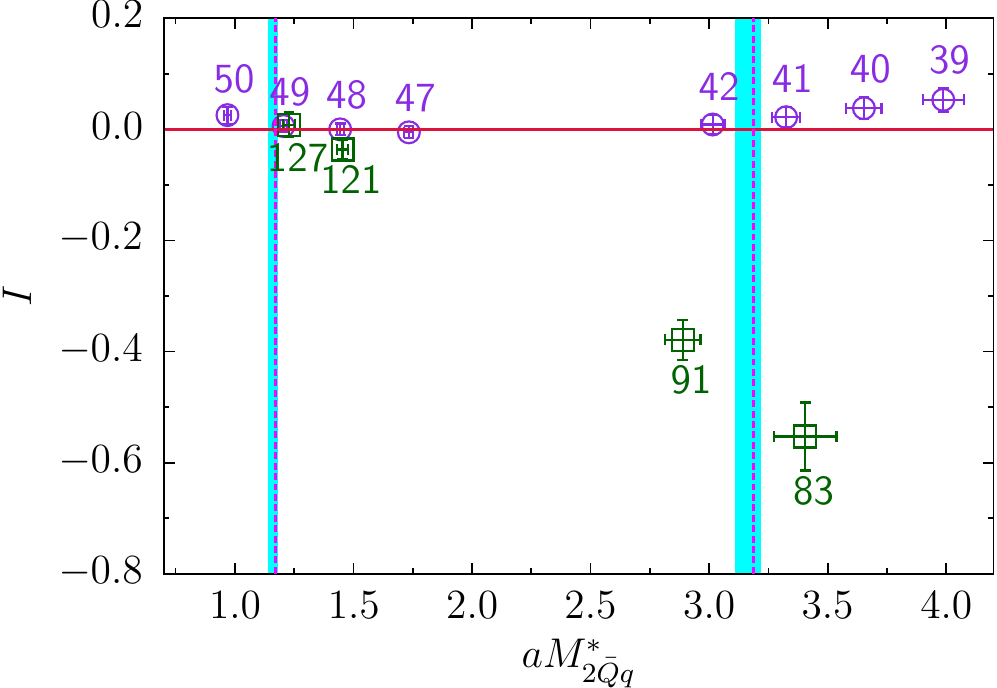}
    }
    \caption{Inconsistency~$I$ for (a)~the spin averaged mass $\wbar{M}=\frac{1}{4}(M+3M^*)$, 
    (b)~pseudoscalar-meson mass, and (c)~vector-meson mass.
    Data labels denote the value of~$\kappa \times 10^3$.
    The purple circles (green squares) represents the OK (Fermilab) action.
    The errors are from the jackknife.
    Vertical lines with bands represent the physical masses from the PDG~\cite{Olive:2016xmw} with 
    experimental and (asymmetric) lattice-spacing errors added in quadrature.
    For the OK action $I$~almost vanishes (cf.\ horizontal red line), but for the Fermilab action it does 
    not.}
    \label{fig:iparam}
\end{figure}

We calculate the inconsistency~$I$ from the pseudoscalar and vector masses presented in
Sec.~\ref{sec:disp}.
The derivation of Eq.~\eqref{eq:NRdeltaB} does not include spin-dependent effects, which should contribute
to $I$ for both pseudoscalar and vector channels.
To separate spin-dependent effects, which are the subject of Sec.~\ref{sec:hfs}, from spin-independent ones,
we form the spin-averaged mass,
\begin{equation}
    \wbar{M}=\frac{1}{4}(M+3M^*),
\end{equation}
writing $M$ ($M^*$) for pseudoscalar (vector) masses.
We calculate the quantity~$I$ with such spin-averaged rest and kinetic masses and plot the result in
Fig.~\ref{fig:iparam-avg}.
For orientation, we also show the physical $B_s$ and $D_s$ masses (for this ensemble) with vertical bands.
We find that $I$ is close to 0 for the OK action even in the $b$-like region, whereas the Fermilab action
leads to a very large deviation~$I\approx-0.6$ from the continuum value, $I=0$, as in Ref.~\cite{Collins}.
This outcome provides good numerical evidence that the improvement of spin-independent effects of the OK
action is realized in practice.
The small~$I$ remaining stems from still higher-dimension kinetic operators of $\order(\lambda^4)$ in HQET,
or $\order(v^7)$ in NRQCD, power counting, which are not addressed by the OK action.
For completeness, Figs.~\ref{fig:iparam-ps} and~(c) show $I$ for the pseudoscalar and vector channels,
demonstrating that spin-dependent effects do not alter the conclusions.

\section{Hyperfine Splittings}
\label{sec:hfs}

The hyperfine splitting is the difference in the masses of the vector and pseudoscalar mesons:
\begin{align}
    \Delta_1 &= M_1^{\ast} - M_1 \label{eq:Delta1}, \\
    \Delta_2 &= M_2^{\ast} - M_2 \label{eq:Delta2}.
\end{align}
From Eq.~\eqref{eq:deltaB}, one has
\begin{equation}
    \Delta_2 - \Delta_1 = \delta{B^{\ast}} - \delta{B} .
\end{equation}
Spin-independent contributions cancel in this binding-energy difference, so the hyperfine difference
$\Delta_2-\Delta_1$ diagnoses the improvement of the spin-dependent $c_3$ and $c_5$ terms of
$\order(\lambda^3)$ in HQET power counting, or $\order(v^6)$ in NRQCD.

\begin{figure}[bp]
    \centering
    \subfigure[~Quarkonium]{
        \label{fig:hfs-hh}
        \includegraphics[width=0.9\columnwidth]{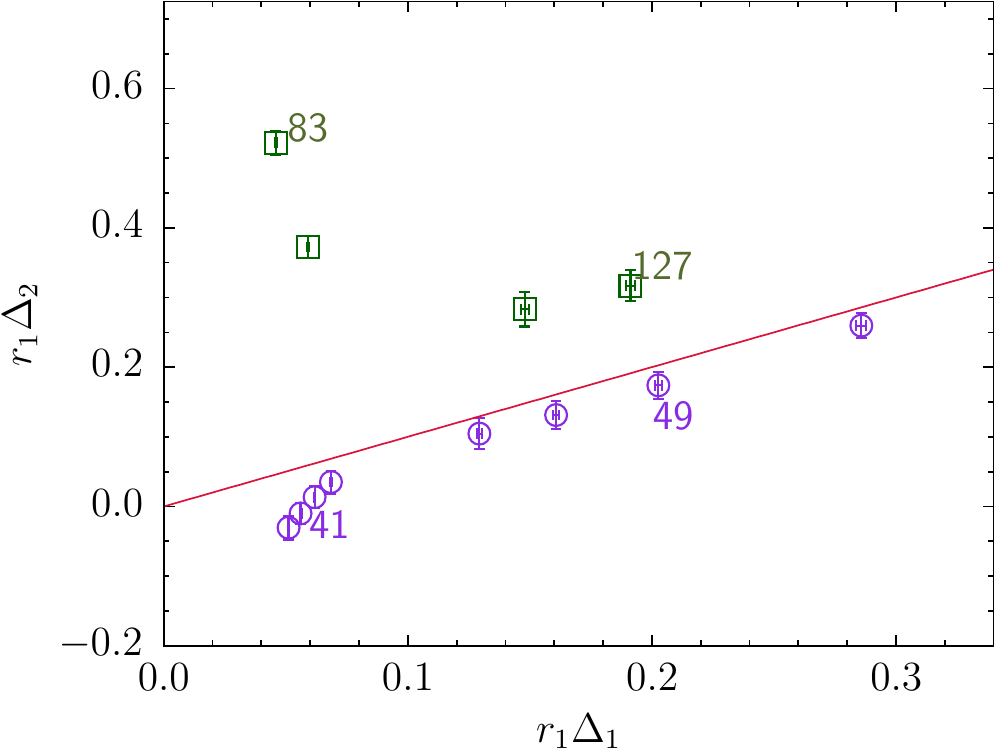}
    }
    \hfill
    \subfigure[~Heavy-light meson]{
        \label{fig:hfs-hl}
        \includegraphics[width=0.9\columnwidth]{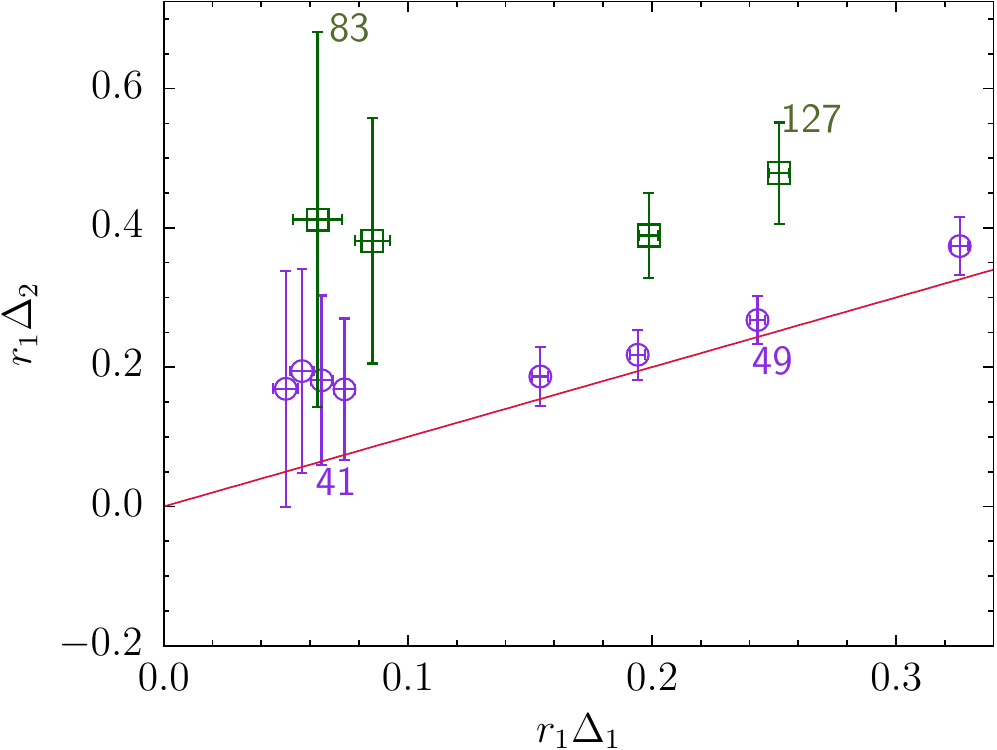}
    }
    \caption{Hyperfine splitting $\Delta_2$ obtained from the kinetic masses vs.\ $\Delta_1$, obtained from 
    the rest masses.
    The square (green) represents the Fermilab action data, and the circle (purple) represents the OK action
    data.
    The labels are $\kappa\times10^3$, corresponding to kinetic masses close to the physical $B_s$ (83, 41) and
    $D_s$ (49, 127) masses, as shown in Fig.~\ref{fig:iparam}.
    The continuum limit is represented by the line (red) $\Delta_2=\Delta_1$.
    Errors are from the jackknife.}
    \label{fig:hfs}
\end{figure}

As one can see in Fig.~\ref{fig:hfs-hh}, the OK action shows clear improvement for quarkonium.
The data points from the OK action lie much closer to the continuum value $\Delta_2 = \Delta_1$ (the red
line) for all simulated values of $\kappa_\text{OK}$; in the charmonium region, they remain
consistent with the continuum line within the error.
The heavy-light results in Fig.~\ref{fig:hfs-hl} also show clear improvement in the region near the $D_s$
mass.
The results with the OK action remain consistent with the continuum limit throughout the $B_s$ mass region,
but the improvement is not yet statistically significant.
Higher statistics would resolve the issue.
All in all, the hyperfine splittings show the improvement from the higher-dimension chromomagnetic
interactions---those with couplings $c_3$ and $c_5$.

For both quarkonia and heavy-light mesons, the hyperfine splitting of the kinetic mass, $\Delta_2$, has a
larger error than that of the rest mass, $\Delta_1$, because the kinetic mass requires correlators with
$\bm{p}\neq\bm{0}$, which are noisier than those with $\bm{p}=\bm{0}$.
As the rest mass $M_1$ and the kinetic mass $M_2$ are determined with smaller error with the OK action than
the Fermilab action (see Sec.~\ref{sec:disp}), the statistical errors shown for the hyperfine
splittings $\Delta_1$ and $\Delta_2$ in Fig.~\ref{fig:hfs} are smaller with the OK action, especially in the
case of $\Delta_2$.

\section{Conclusions}
\label{sec:con}

Our tests of the Fermilab improvement program are based on the two-point correlators for mesons generated
with the Fermilab~\cite{EKM} and OK~\cite{Oktay2008:PhysRevD.78.014504} actions.
Looking ahead to phenomenological applications~\cite{Bailey:2014jga,Bailey:2016wza,Jeong:2016eot}, we have
chosen four hopping parameters for the OK action in each of the $b$- and $c$-quark mass regions.
For the Fermilab action, we have simulated two $b$-like and two $c$-like hopping parameters.
Although the lattice data span the physically interesting regions, the central aim of this paper is to test
the improvement theory of Refs.~\cite{ EKM, Oktay2008:PhysRevD.78.014504} against simulation data,
independent of the phenomenological interpretation of the action's parameters.

We focus on tests of both the spin-independent and spin-dependent terms in the OK action.
The results for the (spin-averaged) quantity $I$, known as the inconsistency~\cite{Collins}, show that the
OK action succeeds in improving the effects that generate the kinetic-mass binding energy.
The hyperfine splitting shows that the OK action significantly improves the higher-dimension spin-dependent,
chromomagnetic effects, at least in quarkonium.
For the heavy-light system, the data show a clear improvement for smaller $c$-like masses, but in the
$b$-like region, large statistical errors prevent us from reaching firm conclusions.

This study has yielded some noteworthy byproducts.
As $am_0$ is reduced, here with fixed $a\approx0.12$~fm, the meson masses $M_1$, $M_2$, and $M_4$ approach
each other, verifying the expectations of the Fermilab and OK actions.
The difference between $M_4$ and $M_2$, for $am_2\not\ll1$, is not much better for the OK action than the
Fermilab action.
By analogy with the inconsistency~$I$~\cite{Collins,Kronfeld}, this feature is probably explained by the
associated binding energy $M_4-m_4$, which stems from higher-dimension effects not improved by the OK
action.
Finally, in the $b$-like region, $am_2\sim3$, the OK action produces statistically more precise results than
the Fermilab action for the heavy-light correlator energies and, hence, the masses $M_2$ and~$M_4$.

As an application of the OK action, a calculation of the form factors for the $B\to D^{(*)}\ell\nu$
semileptonic decays is underway, with the aim of determining the CKM matrix element $|V_{cb}|$.
To achieve the desired sub-percent precision on the relevant form factors, it will be necessary to derive
the analog of the OK action for currents \cite{Bailey:2016wza,Bailey:2014jga}, and to calculate the
renormalization.
Meanwhile some of us are extending the present work to a full-fledged tuning run~\cite{Jeong:2016eot}.

\vspace{1em}
\acknowledgments

J.A.B.\ is supported by the Basic Science Research Program of the National Research Foundation of Korea
(NRF) funded by the Ministry of Education (No.~2015024974).
This project is supported in part by the U.S.\ Department of Energy under grant No.\ DE-FC02-12ER-41879
(C.D.) and the U.S.\ National Science Foundation under grant No.\ PHY10-034278 (C.D.).
A.S.K.\ is supported in part by the German Excellence Initiative and the European Union Seventh Framework
Programme under grant agreement No.~291763 as well as the European Union's Marie Curie COFUND program.
Fermilab is operated by Fermi Research Alliance, LLC, under Contract No.\ DE-AC02-07CH11359 with the United
States Department of Energy.
The research of W.L.\ is supported by the Creative Research Initiatives Program (No.~20160004939) of
the NRF grant funded by the Korean government (MEST).
W.L.\ would like to acknowledge the support from KISTI supercomputing center through the strategic support
program for the supercomputing application research (No.\ KSC-2014-G3-003).
Further computations were carried out on the DAVID GPU clusters at Seoul National University.

\bibliographystyle{apsrev4-1} 
\bibliography{refs} 

\end{document}